# Attempts of Transforming Teacher Practice Through Professional Development

Vera Montalbano[1], Roberto Benedetti[1], Emilio Mariotti[1], Maria Alessandra Mariotti[2] and Antonella Porri[3]

[1]Department of Physics, University of Siena, Siena, Italy
[2]Department of Mathematics and Computer Science "R. Magari", University of Siena, Siena, Italy
[3]Regional Scholastic Office of Tuscany - Arezzo Territorial Area, Arezzo, Italy

Abstract

A difficult challenge in physics education is to design professional development programs for teachers, which can lead to fundamental changes in their practice. We report all activities for physics teachers in the context of the National Plan for Scientific Degrees in Southern Tuscany. Research and practice have shown that physics teaching in school is inadequate. The main consequences are limited achievements in school, decrease of students' interests in learning physics and decrease of enrolments in physics in many countries. In recent years, the decline in enrolments was faced up with the launch of a wide national project addressed to secondary school students and teachers. The active involvement of teachers in the design of laboratories was found to be essential for obtaining actions which were not transitory and entered permanently in classroom practice. We describe some advanced courses in Physics and Mathematics Education realized few years ago and courses designed for a Master in Physics Educational Innovation and Orienting performed jointly by many Italian universities. Other activities are less formal but equally relevant, such as the active involvement of expert, young and in training teachers in designing and implementation of laboratory activities for a summer school of physics. Recently, we developed a workshop for teachers of physics and mathematics on modelling. which continued in an updating course for teachers in which selected topics, named in the same way in both disciplines, were discussed in order to design interdisciplinary learning paths. The purpose is to clarify these topics by using specific tools from physics and mathematics and to outline the similarities and the differences in both contexts. We believe that this activity can be useful for students, which can acquire a profound insight on selected fundamental concepts, and for teacher professional development. We describe teacher reactions and the more significant difficulties we encountered. Finally, we discuss which kind of activity seems more effective.

Correspondence concerning this article should be addressed to Vera Montalbano,
Department of Physics, University of Siena, via Roma 56, 53100 Siena, Italy.
E-mail: montalbano@unisi.it

The decline of interest in studying science is a serious concern for any society in which technology and science are essential in order to achieve an economic prosperity. In the past decades, an impressive decreasing of interest of young people in pursuing scientific careers was observed almost everywhere in the world (Czujko, 2002; Bucchi & Neresini, 2004; National Science Board, 2007; Convert, 2005; Mulvey & Nicholson, 2011). Furthermore, research indicated widespread scientific ignorance in the general populace (Durant & Bauer, 1997; Durant, Evans, & Thomas, 1989; Miller, Pardo, & Niwa, 1997). Then it is essential to educate in sciences as many students as possible to the highest level in the school for obtaining the goal of enhancing the scientific literacy and increasing the number of motivated and talented students enrolling in courses scientific degrees. The quality of teaching plays a crucial role in this context (Osborne, Simon, & Collins, 2003).

During the last decade, we realized several professional development programs for teachers in Physics and Mathematics in order to promote fundamental changes in their practice. Our activities were directed to teachers living in southern Tuscany (provinces of Arezzo, Grosseto and Siena).

In the beginning, we were concerned with teacher training in the Advanced School for Teaching in Secondary School of Tuscany (a 2-years post-graduated course required, at that time, for obtaining teaching qualification in Italy[1]). Since almost all teachers (expert or in training) in our region are not physicists but rather mathematicians or engineers, a strong requirement arose: Put laboratory in the centre of physics education. Designing professional development programs for teachers, which can lead to fundamental changes in the quality of their teaching, was indispensable.

In the following, we present a survey of selected activities that we realized in recent years in order to transforming teacher practice through professional development. Since almost all activities were realized within the National Plan for Scientific Degrees, in the next section we present the plan focusing on the methodological choices which were transposed in the professional development programs for teachers. In the following one, we describe in details a selection of activities, focusing on teacher reactions and the more significant difficulties we encountered. In the last section, we discuss the obtained results and which type of activities appears more effective. In particular, the context in which the activity is realized seems to assume an important role.

## Actions for Scientific Degrees in Italy

In recent decades it has been detected almost everywhere a consistent decrease of graduates in science disciplines. The situation in Italy was dramatic; enrolments in basic sciences are more than halved in few years (Vittorio, 2010). Therefore, the Ministry of Education and Scientific Research promoted a wide project in order to reverse this trend. Starting at the end of 2005, the project was named Scientific Degree Project (Progetto nazionale per le Lauree Scientifiche, i.e. PLS) and was financed for four years. During this period and at the end of the project, a large monitoring of all activities was realized in order to identify what actions were more effective and incisive in contrasting decreasing of scientific vocations. The main actions of PLS were professional development for teachers and orienting for students essentially by means of laboratory activities (see for example Sassi, Chiefari, Lombardi, & Testa, 2012). In 2009 it was launched the National Plan for Science Degree (same acronym PLS) where some of the most effective methodological aspects were emphasized in new guidelines (PLS website; Vittorio, 2010).

---

[1] The Advanced Schools for Teaching in Secondary Schools became the only way for obtaining teaching qualification for teaching in schools of first and second degree. Activated in the Academic Year 1999-2000 with the cycle I, they were finally closed in the Academic Year 2008-2009 at the end of the ninth cycle. They were organized at inter-regional level, with closed number access (set each year by the Ministry of University and Research). Starting with the Academic Year 2012-2013, the teaching qualification can be achieved by following an annual course named Formative Active Training.

**Scientific Degree Project 2006 - 2009**

The project originated from a collaboration of the Ministry of Education and Scientific Research, the National Conference of Deans of Science and Technology and Confindustria, the main organization representing Italian manufacturing and services companies, and was designed with the initial motivation to increase the number of students on degree courses in Chemistry, Physics, Mathematics and Science of Materials.

The project focused on three main objectives (Vittorio, 2010):

- improving knowledge and awareness of science in secondary school, offering students in the last three years of school to participate in stimulating and engaging curricular and extra curricular activities in laboratory;
- starting a process of professional development of science teachers in service in the secondary school from joint work between School and University for the design, implementation, documentation and evaluation of the laboratories mentioned above;
- promote alignment and optimization of training from University and School for the working world.

The action for student integrated with training for teachers was made through more than 100 sub-projects under the responsibility of local referents, located in 33 universities, became 38 in the last period, spread all over the country and organized into four areas of national projects (PLS website; Vittorio 2010).

**National Plan for Science Degree 2010 - 2012**

The plan maintained the same purposes of increasing the enrolment in science degrees to which was added the necessity to revise the content and methods of teaching and learning of science in all grades of school, taking into account the new national guidelines for first and second cycle contained in the recent Italian reform of the educational system.

**Strategy and Methodologies**

In order to achieve the above purposes, the Plan pursued the strengthening and the practical realization of the main ideas that were effective in trials 2005-2009 (PLS website):

- orientation does not conceive how a teaching path given to student, but as an action that the student is doing, from meaningful activities that allow to compare problems, issues and ideas of science;
- designing the training of teachers in service by involving teachers in solving concrete problems, developing design and implementation of educational activities and through comparison with peers and experts;
- pursuing and achieving at the same time the student orientation and training for teachers through the planning and joint implementation by school teachers and university laboratories for students, thus developing relations between the school system and the University;

Furthermore, a new idea was added: consciously connecting the activities of the Plan with the innovation of curricula and teaching methods adopted in schools, and other contents and methods of teacher training (initial and in-service), for the first and second cycle. Thus, the main road consisted in considering laboratory as a method, not as a place. Students must have become the main character of learning and joint planning by teachers and university was a mandatory step.

More attention to laboratory is required and different types of laboratory PLS can be proposed: Laboratories which approach the discipline and develop vocations, Self-assessment laboratories for improving the standard required by graduate courses, Deepening laboratory for motivated and talented students (for a survey on PLS laboratories realized locally see Montalbano, 2012).

## Professional Development Programs in Southern Tuscany

Since 2000, the Department of Physics is engaged in many activities with the purpose of enhancing cultural knowledge and skills of training and in-service teachers. First advanced course started in order to satisfy a request from young teachers for further examples of designing learning path in physics and mathematics after their teacher qualification in the Advanced School for Teaching in Secondary School of Tuscany. The experience was positive, the purpose and the methodology were very close to the PLS and therefore we decided to propose this course for the project. The social context, and national legislation continuously changes with respect to teaching, required that new proposals were designed almost every one or two years. In the following, these activities are briefly described focusing on the typology of participants, the effectiveness of the activity, encountered difficulties and achievements, and so on. The evaluation arose from observations during the activity and discussions performed at the end by all people, faculty staff and expert teachers, which were involved in the professional development program.

### Advanced Courses for Teachers of Physics and Mathematics

Starting in 2004, we organized an annual advanced course for professional development of qualified teachers, entitled *Learning paths in Physics and Mathematics: Models, Experimental checks, Statistics*. Teachers were engaged in designing learning paths with particular attention to relationship between physics and mathematics.

**Activity description.** The main features of this program were a long period activity (100-125 hr spread over several months), compulsory attendance, a score for teacher ranking available after final examination, a huge period dedicated to activities in laboratory, data analysis and discussion on educational aspects. The activities were divided in two courses: Society and education (25-35 hr) and Educational Research Laboratory (75-90 hr).

*Society and education.* We presented educational resources that can be found outside the school (e.g. science museums, interactive science exhibitions), conferences on history of physics or mathematics, cultural reflections that can enrich the quality of teaching. When the conference was of more general interest it was open to public.

*Educational research laboratory.* Experts from school, university and research institutions presented innovative tools for education in three topics selected from the school curriculum. Small groups of teachers elaborated these topics, by designing and implementing learning paths in the laboratory and analysing experimental data. At the end, they produced materials in order to discuss some learning paths in the final examination. Some examples of proposed educational research laboratories are the following: Modelling and approximating in mathematics and physics, Wave phenomena, Educational project on modern physics, Introduction to relativity, Waves and oscillations: a methodological reflection, From electrical charge to Joule effect, Waves: from sound to light, From special to general relativity.

**Participants.** We proposed the advanced course for four academic years. Enrolments for each year are showed in Table 1. Few participants had a permanent position but all were teaching at school. The decreasing of enrolments corresponds to a change in rules for attributing score for teacher ranking and the proliferation of online courses that gave the same score with fewer costs and efforts for teachers.

Table 1

*Qualified teachers enrolled in Advanced Course Learning paths in Physics and Mathematics*

| Academic year | 2004/05 | 2005/06 | 2006/07 | 2007/08 |
|---|---|---|---|---|
| Participants with permanent position | 2 | 3 | 4 | 0 |
| Participants without permanent position | 15 | 16 | 10 | 3 |
| Total enrolments | 17 | 19 | 14 | 3 |

**Activity analysis.** The goal of the advanced course was achieved. Participants selected, within all teaching suggestions, one or two ideas that were developed by preparing educational materials ready for the use in classroom and laboratory. Few teachers were able to test in their classes the learning path and evaluate its effectiveness. Moreover, we identified the following critical points:

- the course was very intensive for teachers;

- the most part of the participants were young teachers with no permanent position, thus with no possibility of engaging in long term experimentation or stable collaboration with colleagues of their schools. Thus, the possible impact of the intervention on school practice is limited.

- every year it became more difficult to organize the course (University taxes became too high);

- online course organized by private institutions gave the same score for teacher ranking with fewer costs, time engagement and effort for teachers.

**Training in Pigelleto's Summer School of Physics**

Since 2005, forty students from high school are selected to attend a full immersion summer school of physics in the Pigelleto Natural Reserve, on the south east side of Mount

Amiata in the province of Siena (Benedetti, Mariotti, Montalbano, & Porri, 2011). Perhaps, this is the most successful activity that we have ever realized in PLS for orienting students towards physics.

Starting from the third edition of the school, teachers enrolled in Advanced School for Teaching in Secondary School of Tuscany were invited for a stage at the summer school. After the closure of the Advanced School for Teaching in Secondary School, we still continue to invite young teachers for a full-immersion stage.

**Activity description.** The school begins usually in early September and lasts for four days. The 2011 edition was entitled Thousand and one energy: from sun to Fukushima; some previous editions were Light, colour, sky: how and why we see the world (2006), Store, convert, save, transfer, measure energy, and more…(2007), The achievements of modern physics (2009), Exploring the physics of materials (2010). Topics are chosen so that students are involved in activities rarely pursued in high school, relationships with society are outlined and discussed. The students are selected by their teachers in the network of schools involved in the National Plan for Science Degree [8]. In the morning, we propose lessons in which the necessary background for the following activities in laboratory is given. In the afternoon, small groups of students from different schools and classes are engaged in laboratories where are forced to take an active role. All groups are supported by one or two teachers that are available to discuss any idea. Usually we propose different laboratories and for each one the group of participants is asked to prepare a brief presentation for sharing with other students what they have learned. After dinner, an evening of astronomical observation of the sky is usually expected. If it is cloudy, a problem solving evening is proposed. In designing activities, we pay attention to several aspects that can render this action more effective both for teachers that for students, for example the main topic is related to all activities and must not be trivial. In order to have the best collaboration, students' groups are inhomogeneous and formed by following the teachers suggestions. When it is possible, laboratories are made with poor materials or educational devices provided by a school in such way that teachers can duplicate easily the lab in their context. In order to focus student and teachers attention to physics as an experimental science, almost all laboratories lead to at least one measure and its error valuation. Last but not least, methodologies are discussed and selected with the teachers involvement and usually in laboratory an expert and a young teacher are engaged in order to improve teacher practise. Let us give some examples of laboratory: Measurement of Plank's constant by a LED, Measure of the speed of rotation of a star, How it works: Stirling machine, solar oven, coffeepot, Measurement of the mechanical equivalent of heat, Electromagnetic induction and energy dissipation, Radioactivity background vs. small Uranium sources, Photoelectric effect.

**Participants.** Usually there are about 4-5 expert teachers and 5-7 young teachers. In the last edition some young teacher acquired enough experience to become expert.

**Activity analysis.** This informal training seems to be very effective. Young teachers benefit from direct experience in laboratory with expert teachers. Participants sometimes seem to be inspired by the school's laboratories and some activities entered in practise. On the other side, we identified the following critical points:

- the summer school was very intensive for teachers and often they are obliged to interrupt the stage for going back to school service;

- young teachers (almost all with no permanent position) can be called for a temporary job during the school;

- teachers in-service almost never participate to the summer school (they usually prefer not to accompany their student).

**Master in Physics Educational Innovation and Orienting**

In recent years, a national master designed for qualified teachers has been organized by University of Udine. We joined the master's third edition where nineteen universities all over the country collaborated for giving courses in laboratory, often focusing on laboratories performed within the National Plan (Stefanel et al., 2012)

**Activity description.** Participants can choose between a variety of courses. Many courses are online, but laboratories requiring presence must be taken in one of the universities. After few introductory lessons in which methodological and disciplinary aspects were treated usually by discussing as example a PLS laboratory, teachers were asked to design a specific learning path. When it was possible, paths were then tested in class. The courses given by University of Siena have been the following: Modern physics, Advanced physics experiments, Spectroscopy laboratory, Look at the universe: astrophysical observations and measurements, Waves and oscillations, Phenomenology of sound, Paths on thermal phenomena, Paths on electromagnetism, Exploratory lab on superconductivity. The master allows two years of attendance, a final thesis must be discussed as final examination. Teachers can attend full activities of master or, taking fewer courses, attend an advanced course or finally attend only few courses that can be certified.

**Participants.** Teachers all over the country enrolled. Locally, we had one teacher for the master and two for single courses.

**Activity analysis.** The master raised a higher standard for the professional development because many activities inspired physics educational research in teachers (see for example Di Renzone, Frati, & Montalbano, 2011), especially on topics rarely touched in class or in laboratory. As in the case of advanced courses, we identified the following critical points:

- the master was too intensive for in-service teachers;

- only few young teachers (with no permanent position) enrolled; therefore, there was no possibility for a change in practice.

**Laboratory on Modelling**

Since in 2009 also the Department of Mathematics and Computer Science of University of Siena launches its plan, a workshop for teachers of physics and mathematics on modelling was performed in collaboration with Physics Plan.

Mathematics is frequently introduced as a set of tools and techniques for modelling real world situations and solve real-world problems (for a characterization of modelling in mathematics education, see 122 in Niss, 2003). Thus, the relationship between mathematics and other sciences, in particular with physics, becomes purely instrumental and user tends to learn how to utilize it without worrying about understanding how and why a specific mathematical tool is useful for achieving a goal.

Also in physics, modelling process is central and physicists tend to use modelling as a research tool. For a brief survey of models in physics and physics education see for example p. 256 in Angell, Kind, Henriksen, and Guttersrud (2008) and for a wider overview in contemporary science education see GIREP (2006).

If a learning path describes phenomena without accounting for the process that led to their quantitative description, the modelling process remains hidden. Sometimes this process has a long and complex history, which shows the presence of epistemological obstacles that required time and intellectual development to be overcome. We believe that the same problem might arise for students and it can be an opportunity for a deeper learning process. Thus, not taking into account the complexity of the modelling process that is the origin, the phenomenon and its mathematical description seem to be as distinct features, artificially glued, and therefore often fail to interact.

The aim of this laboratory was to try to re-establish the relationship of synergy between the description/explanation of the experience and the mathematical model to be taught, presenting students work activities (explorations and problems) centered on the process of modelling: (a) starting from a rough problem, (b) the emergence of hypotheses, (c) varying quantities with the purpose of obtaining all relevant relationships between variables.

**Activity description.** The laboratory was performed with teachers, mathematicians and physicists, and we encountered every month for a total of 20 hr. We started by proposing selected topics, that in the curricula appear named in the same way in both disciplines. A first discussion was devoted to identify the particular perspective from which the 'same' topic was considered in the two disciplines. The final objective was the interdisciplinary design of learning paths that could exploit the different perspectives to make student construct unified and richer conceptions. The purpose was to make a specific conception emerge from a laboratory experience and then find a formalization in physics and mathematics: outlining similarities and differences in both contexts. Rough problems are proposed in order to inspire learning paths in which a modelling process can arise and develop in the different disciplinary fields. Though we believe that this kind of activity can be proposed only for specific topics, we believe that it would be useful for students to have the opportunity of experiencing, at least in some cases, how the modelling process can develop, and consequently acquire a profound insight on some fundamental concepts. The experience of designing this type of didactic path has been of great value for teacher professional development. The introductory discussions, to which physicists and mathematicians contributed bringing their own different perspectives, faced the teachers with epistemological and cognitive problems that they had never foreseen.

**Participants.** In the beginning, there were about 10 teachers (2 with a permanent position in school), but at the end only 4 remained.

**Activity analysis.** The activity was very stimulating for the participants but materials for the practice in class was slowly prepared after long discussions where the different cultural backgrounds of physicists and mathematicians involved emerged clearly. Only a couple of proposals were tested successfully in classroom. Because of the inter-disciplinary character of these kind of activities, it was difficult to experiment in the classroom, actually there are a few opportunity in this field in our schools. Moreover, we identified the following critical points, which we already find in other cases:

- meetings were too intensive for in-service teachers and there was an objective difficulty in conciliating the schedule of meetings with all teachers duties;

- again the most were young teachers (with no permanent position) limiting the possibility of sharing this experience with other colleagues. However, there was a great attention from some teachers in permanent position.

Thus, we decided to continue this experience the following year in an updating course for teachers in a school.

**Updating Courses for Teachers In-Service**.

**Overview.** Updating is compulsory for teachers in-service (6 hours every year), therefore there is a certain interest for teachers to follow updating courses that are really useful for their teaching practise. Different schools have different teams of teachers and different ways of perform the updating. Anyway, in these years we have tried to propose interesting and useful updating courses in order to improve teaching of physics and mathematics. The courses were often tailored on the specific context of the school where they were realized.

**The nature of light: from classical physics to quantum physics.** This course was performed in a scientific high school (Liceo Scientifico Statale *Redi* in Arezzo) in 2008/2009.

*Activity description*. For 10 afternoons we met participants in the physics laboratory of the school for a total of 40 hr. The first hour was dedicated to educational considerations on selected topics of physics, the rest of the time teachers were involved in performing experiments by using the facilities of the laboratory in the school.

*Participants.* There were 17 teachers enrolled, 13 of them regularly attended. The course was promoted in all schools in Arezzo and surroundings but the most of participants were in-service in that scientific high school.

*Activity analysis*. The activity worked very well. Teachers were all in-service and were strongly motivated in learning how to use their lab's facilities because of the forthcoming reform of secondary school. Moreover, we got a strong school's management support. Another important consideration is that the course was realized in a peculiar school, with very well furnished physics laboratory, but only few teachers were able to use this opportunity in their practice properly because their background was lacking in physics practise.

**Course on physics laboratory in the new high schools.** This course was performed in Liceo Scientifico Statale *Redi* in Arezzo in 2009/2010.

*Activity description*. For 8 afternoons we met participants in the school for a total of 24 hr. Teachers were interested in designing a meaningful curriculum for physics in a reformed school. They produced a detailed set of learning paths which covered the physics program for the first year of the reformed scientific high school.

*Participants.* There were 11 teachers enrolled that regularly attended the course.

*Activity analysis.* The activity was very satisfactory. Teachers were all in-service and are strongly motivated. There was again a strong school's management support. Another aspect is that all teachers were in-service in that peculiar school.

**Course on physics laboratory.** The course was performed in a scientific high school (Liceo Scientifico Statale *Volta* in Colle Val d'Elsa, Siena) in 2009/2010, in order to satisfy a request from teachers in-service there.

*Activity description*. For 3 afternoons we met participants in the school for a total of 6 hr. Teachers were interested in performing some activity in their physics lab (few years ago the lab caught fire and all the surviving material remained unused). One of the authors (E. M.) went to find out what they had in their closets and then presented some learning paths in which teachers could use their supplies.

*Participants.* There were 6 teachers enrolled that had regularly attended the course.

*Activity analysis*. The activity was unsatisfactory. Teachers interested but not active They concerned only to obtain some useful hint for physics lab. Moreover, there was a weak school's management support and all teachers were in-service in that school.

**Mathematics and physics teaching in the reformed school.** The course took place in Scientific High School (Liceo Scientifico Statale *Galilei* in Siena) in 2011/2012, in order to satisfy a request from teachers in Laboratory on modelling.

*Activity description*. For 10 afternoons we met participants in the school for a total of 30 hr. The methodology was introduced by some examples belonging from material elaborated in the Laboratory of modelling the previous year. Afterward, selected rough problems were proposed and participants realized a laboratory activity. The next step was to discuss together if the activities were effective or could be modified in order to achieve the goal of realizing an active process of modelling.

*Participants.* Despite 19 teachers enrolled, only 8 have regularly attended the course.

*Activity analysis.* The activity was unsatisfactory. Many teachers were interested only in obtaining some useful hint or ready recipe for physics lab, but they strongly opposed to challenge their way of teaching. Moreover, school's management support was totally absent (only 2 participants were in-service in the school). The only positive aspect was that we started with few interested teachers a fully designing of interdisciplinary activities for the first year of the reformed scientific high school and for a technical school.

## Discussion and Conclusions

Professional development is indispensable in order to obtain a real and permanent improvement of physics teaching. An active engagement of teachers in the design and implementation of learning paths closely related to laboratory activities is an essential step for trying to change the current teaching practice. Another key point is to experiment in the classroom or with groups of students the activities designed in order to clarify which ones are more effective and to increase the experience and confidence that teachers have in laboratory practices. The whole process from designing a didactic path to implementing it in a class is necessary for a teacher to develop the autonomy for innovation, based of a solid disciplinary background and a critical attitude towards educational issues.

Unfortunately, we found very difficult to involve actively in-service teachers, especially if they had a permanent position in school. At the origin of this difficulty, there is not only the intense engagement in school duties, but also a difficulty of changing well settled didactical habits, in spite of a general dissatisfaction in terms of learning achievements. Young teachers (often with no permanent positions) were more available to innovation and more disposed to invest their time to enhance their teaching skills. However, sometime the impact with the reality of the school makes them give up.

Updating courses are almost the only way to achieve teachers in-service (better if held in their school) and can be designed taking in account of the context. In this case, strong school's management support is crucial in order to obtaining a wider teachers' involvement.

### Acknowledgement

This work is based on activities and experiences which were realized within the National Plan for Science Degree supported by Italian Ministry of Education, University and Research. The authors would like to thank the technical staff of Department of Physics and of Liceo Scientifico Statale *Redi* for the availability and continuous support always essential for maintaining fully operational physics laboratories.

References

Angell, C., Kind, P. M., Henriksen, E. K., & Guttersrud, Ø., (2008), An empirical-mathematical modelling approach to upper secondary physics, *Physics Education*, *43* (3), 256-264.doi:10.1088/0031-9120/43/3/001

Benedetti, R., Mariotti, E., Montalbano, V., & Porri A., (2011), Active and cooperative learning paths in the Pigelleto's Summer School of Physics, *Twelfth International Symposium Frontiers of Fundamental Physics (FFP12)*, Udine 21-23 November 2011, arXiv:1201.1333v1 [physics.ed-ph]

Bucchi, M., & Neresini, F. (2004), The scientific vocation crisis and its explanation, *Observa – Science in Society*, 1-41, http://chemistry.pixel-online.org/database_review.php?id_doc=70

Czujko, R. (2002), Enrollments and Faculty in Physics, pp. 1-11 http://www.aip.org/statistics/trends/reports/june9talk.pdf, accessed 2012 March.

Convert, B. (2005) Europe and the Crisis in Scientific Vocations, *European Journal of Education*, *40*(4), 361-366.

Di Renzone, S., Frati, S., & Montalbano, V., (2011), Disciplinary knots and learning problems in waves physics, *Twelfth International Symposium Frontiers of Fundamental Physics (FFP12)*, Udine 21-23 November 2011, arXiv:1201.3008v1 [physics.ed-ph]

Durant, J., & Bauer (1997). Public understanding of science: the 1996 survey. Paper presented at a seminar at the Royal Society, 8 December 1997.

Durant, J. R., Evans, G. A., & Thomas, G. P. (1989). The public understanding of science. *Nature, 340*, 11–14.

GIREP (2006), Modelling in Physics and Physics Education Conference (Amsterdam NL 20-25 August 2006), http://www.girep2006.nl/, accessed 2012 November.

National Science Board, (2007) The Message of the 2004 S & E Indicators: An Emerging and Critical Problem of the Science and Engineering Labor Force , 1, http://www.nsf.gov/statistics/nsb0407/, accessed 2012 March.

Miller, J. D., Pardo, R., & and Niwa, F. (1997). Public perceptions of science and technology: a comparative study of the European Union, the United States, Japan, and Canada (Bilbao: BBV Foundation).

Mulvey P. J., & Nicholson, S. (2008). Physics Enrollments: Results from the 2008 Survey of Enrollments and Degrees, AIP Focus On, 2011, 1-10, http://eric.ed.gov/PDFS/ED517370.pdf, accessed 2012 March.

Montalbano, V., (2012), Fostering Student Enrollment in Basic Sciences: the Case of Southern Tuscany, in *Proceedings of The 3rd International Multi-Conference on Complexity, Informatics and Cybernetics*: IMCIC 2012, ed. N. Callaos et al, 279, arXiv:1206.4261 [physics.ed-ph].

Niss, M. A. (2003). Mathematical competencies and the learning of mathematics: the Danish KOM project. Gagatsis, A., & Papastavridis, S. (Eds.), In: *3rd Mediterranean Conference on Mathematical Education - Athens, Hellas 3-4-5 January 2003.* (pp. 116-124). Athen: Hellenic Mathematical Society.project, http://w3.msi.vxu.se/users/hso/aaa_niss.pdf, accessed 2012 November.

Osborne, J., Simon, S., & Collins, S. (2003): Attitudes towards science: A review of the literature and its implications, *International Journal of Science Education, 25*(9), 1049–1079.doi:10.1080/0950069032000032199

PLS website, http://www.progettolaureescientifiche.eu/, accessed 2012 March.

Sassi, E., Chiefari, G., Lombardi, S., & Testa, I. (2012), Improving scientific knowledge and laboratory skills of secondary school students: the Italian Plan "Scientific Degrees", *The World Conference on Physics Education (WCPE),* July 1-6, 2012, Istanbul, Poster Session Strand 3: Learning Physics Concepts, P2.G02.03

Stefanel, A., Michelini, M., Altamore, A., Bochicchio, M., Bonanno, A., De Ambrosis, A., … Stella, R. (2012), The Italian project IDIFO3 (Innovazione Didattica in Fisica e Orientamento 3), *The World Conference on Physics Education (WCPE),* July 1-6, 2012, Istanbul, Poster Session Strand 6: Secondary School Physics, P2.G04.06

Vittorio N. (2010), Il Progetto Lauree Scientifiche 2005 – 2009. Sintesi delle attività e dei

risultati a livello nazionale, Workshop Vetrina PLS: da Progetto a Piano, Siena 4 Nov 2010, http://www.unisi.it/fisica/laureescient/vetrina/20101104_vittorio_pls.pdf, accessed 2012 March.